\documentclass{article}
\usepackage[a4paper, top=1.2in, bottom=1.2in, left=1.3in, right=1.3in]{geometry}

\usepackage{hyperref}
\hypersetup{
     colorlinks=true,
     linkcolor=blue,
     anchorcolor=black,
     citecolor=blue,
     filecolor=black,      
     urlcolor=black,
}
\usepackage{amssymb, amsmath, amsthm}
\usepackage{graphicx}
\usepackage{xcolor}
\usepackage{orcidlink}
\usepackage{authblk} % for affiliations
\usepackage{url}
\usepackage{units}

\usepackage{ulem}
\usepackage{xparse}
\usepackage{amsthm}
\newtheorem{proposition}{Proposition}
\usepackage{orcidlink}

\newcommand{\acknowledgement}{\subsection*{Acknowledgement}}

\title{Capturing the Macroscopic Behaviour of Molecular Dynamics with Membership Functions}

\author[1,2]{A. Sikorski\,\orcidlink{0000-0001-9051-650X}}
\author[1]{R. J. Rabben\,\orcidlink{0000-0002-8048-7554}}
\author[1]{S. Chewle\,\orcidlink{0000-0002-1428-349X}}
\author[1]{M. Weber\,\orcidlink{0000-0003-3939-410X}}

\affil[1]{Zuse Institute Berlin}
\affil[2]{Free University of Berlin}

\date{07.06.2024\footnote{Submitted to Proceedings of the Thematic Einstein Semester ``Mathematical Optimization for Machine Learning'', de Gruyter}}

\begin{document}

\maketitle

\newcommand{\mht}{t_\text{mh}}
\newcommand{\infgen}{\mathcal{L}}
\newcommand{\koop}{\mathcal{K}}
\newcommand{\koopt}{\mathcal{K}^\tau}
\newcommand{\reacdir}{\boldsymbol{r}}
\newcommand{\statex}{{\boldsymbol{x}}}

\newcommand{\corrected}{\correctedfinal}
\NewDocumentCommand{\correctedmarkup}{m O{}}{\textcolor{red}{\IfValueT{#2}{\sout{#2} }#1}}
\NewDocumentCommand{\correctedfinal}{m O{}}{#1}

\begin{abstract}
%Markov processes are often used to derive macroscopic quantities (like kinetic quantities) from microscopic dynamics simulations. In our analysis, macro states are defined as functions $\chi:\Omega\rightarrow [0,1]$ which assign  membership values $\chi(\boldsymbol{x})$ to each micro state $\boldsymbol{x}\in \Omega$. We will show that this assignment not only has advantages in terms of consistency (micro vs macro) but also allows for picking a time-determining path out of simulation data of the process. Our illustrative analysis of a molecular dynamics simulation provides such a transition path from an inactive to an active $\mu$-opioid receptor.

Markov processes serve as foundational models in many scientific disciplines, such as molecular dynamics, and their simulation forms a common basis for analysis.
While simulations produce useful trajectories, obtaining macroscopic information directly from microstate data presents significant challenges. This paper addresses this gap by introducing the concept of membership functions being the macrostates themselves.
We derive equations for the holding times of these macrostates and demonstrate their consistency with the classical definition.
Furthermore, we discuss the application of the ISOKANN method for learning these quantities from simulation data.
In addition, we present a novel method for extracting transition paths \corrected{from simulations} based on the ISOKANN results and demonstrate its efficacy by applying it to simulations of the $\mu$-opioid receptor.
With this approach we provide a new perspective on \corrected{the analysis of }[analyzing the] macroscopic behaviour of Markov systems.

%%    We would not ask a quantum chemist where exactly the electron is and what its momentum is.  Thus, the same kind of indeterminacy could apply for the clustering of states of a Markov process in order to arrive at its macro states. ISOKANN will be presented as a tool to generate such a kind of clustering of samples from molecular dynamics simulations. It will turn out that indeterminacy provides a way to  which might be impossible with ``crisp clustering''. {\bf  gefällt mir noch nicht MW}  

\end{abstract}

\section{Introduction}
\label{sec:einleitung}
Many physical processes are best understood as autonomous Markov processes (e.g., \cite{Norris.1998}).
%Many types of physical processes are regarded as Markov processes (e.g., in \cite{Norris.1998}).
A common mathematical model for them are stochastic differential equations which allow to predict the probability for the future evolution of the system. Markovianity means, that this probability is just depending on the initial condition of the system. Solving the initial value problem is referred to as {\em simulation of the system}. The states of the generated trajectory are denoted as the {\em micro states} of the system.

As an illustration, consider the field of biomolecular system simulation \cite{Mura_2014}. Here a common mechanism being studied is the transformation of an inactive micro state of a receptor protein into an active state (by accordingly changing the coordinates of its atoms).
%To give an example: The simulation of biomolecular systems is well studied \cite{Mura_2014}. One possible process under investigation is a receptor protein which is in an inactive micro state and transforms into an active state (by changing the 3D-coordinates of its atoms).
The process can be mathematically modeled by the overdamped Langevin dynamics, a stochastic differential equation for the 3D coordinates of each atom in the protein and the solvent \cite{Leimkuhler.2015}.
Typically these are analysed by simulation of the model, resulting in trajectories in a high-dimensional space.
%Mathematically, the overdamped Langevin dynamics is an often used stochastic differential equation applied to the 3D-coordinates of all atoms of the protein and the solvent to model this process \cite{Leimkuhler.2015}. Simulation generates a trajectory in a high-dimensional space.

However, often, the main focus of interest lies on the {\em macro state behaviour} of the system. 
%However, what we really want to  extract from this simulation is the {\em macro state behaviour}.
A possible question to answer would be: What is the mean first passage time for an inactive receptor protein to become active? The two macro states in this regard are denoted as ``inactive'' versus ``active''. At first glance, macro states $S$ are subsets of the set of micro states $\Omega$. In general, when given a starting set $S\subset\Omega$ of the system we want to know, how long on average the process stays in this set \cite{ernst}. How long does a trajectory starting in the inactive macro state remains there before leaving it, i.e. switching to the active macro state? The answer is given by the mean holding time $\mht(\boldsymbol{x})$ defined by the integral:
\[
\mht(\boldsymbol{x}) = \int\limits_{0}^\infty p_S(\boldsymbol{x},t) \,dt. 
\]
where $\boldsymbol{x}$ denotes the micro state of the system (the starting point ${x}(0)=\boldsymbol{x}$ of the trajectory) and the function $p_S(\boldsymbol{x},t)$ denotes the probability that a trajectory is still in set $S$ and has never left it during the whole time interval $[0,t]$.
\corrected{Formally $p_S(\boldsymbol{x},t) = P_{\boldsymbol{x}} (T_S>t)$ is the holding probability of $S$ conditioned on starting at the micro state $\boldsymbol{x}$ with $T_S$ being the first exit time of $S$. $\mht$ (also called mean first exit time) is then given by 
$ \mht(\boldsymbol{x}) = \mathbb{E}_{\boldsymbol{x}}[T_S]$.}

If $\mht(\boldsymbol{x})$ denotes the expected time until the system reacts, then
\[
\boldsymbol{r}(\boldsymbol{x})=-\nabla \mht(\boldsymbol{x}) = -\nabla \int\limits_{0}^\infty p_S(\boldsymbol{x},t) \,dt
\]
points into the direction where this time decreases the most. This can be seen as the micro-state-dependent reaction path direction $\boldsymbol{r}(\boldsymbol{x})$. 

There is a conceptual problem now. We want to know the mean holding time for the macro state and not for every single micro state, but $\mht$ is a function of the micro states -- the mean holding time in $S$ depends on the starting point. 
Speaking of the ``mean holding time of the macro state $S$'' would only make  sense if it was independent of the micro-states position inside that macro state, i.e. \corrected{if there was a set $S\subset \Omega$ and an exit rate $c_1>0$} allowing for a separation of the type 
\begin{equation}
\label{eq:holdprobclassic}
    p_S(\boldsymbol{x},t) = 1\!\!1_S(\boldsymbol{x})\, \mathrm{e}^{-c_1 t},
\end{equation}

where $1\!\!1_S$ is the indicator function of the set $S$. \corrected{In Markov State Modelling it is often asked for a decomposition of the state space $\Omega$ into subsets $S$ which can be identified as ``macro states'' \cite{Prinz}. The separation approach (\ref{eq:holdprobclassic}) is a prerequisite for Markovianity of the macro state $S$ and is approximately valid, if ``leaving the set $S$'' is a very rare event compared to the mixing within $S$.} The constant $c_1>0$ corresponds to the exit rate of $S$, because in this case $\mht(\boldsymbol{x})=1\!\!1_S(\boldsymbol{x}) \, \frac{1}{c_1}$ is independent \corrected{of} the choice of the starting point in $S$.  This decomposition, however, is not possible in general and corresponds to instantaneous transitions which do not provide reaction paths inside $S$. 

Classically the computation of (micro-state) mean holding times in the case of an $S$-based theory is provided by solving a partial differential equation  \cite{pavliotis}:
\[
\infgen \mht(\boldsymbol{x})= -1,
\]
for all $\boldsymbol{x}\in S$ with the boundary condition $\mht(\boldsymbol{x})=0$ for all $\boldsymbol{x}\not\in S$. 
In this equation the differential operator $\infgen$ is the infinitesimal generator of the Koopmann operator of an autonomous Markov process \corrected{and the equation essentially prescribes the mean holding time to decrease by one time unit per time unit until hitting the boundary.} If the decomposition $\mht(\boldsymbol{x})=1\!\!1_S(\boldsymbol{x}) \, \frac{1}{c_1}$ \corrected{were} valid, then this equation would read 
\begin{equation}
\label{eq:tmh}
\infgen 1\!\!1_S(\boldsymbol{x})= -c_1 1\!\!1_S(\boldsymbol{x}). 
\end{equation}
\corrected{
Under the assumption of an ergodic system (e.g. for non-degenerate diffusion) the only set-based solutions are}[
This \corrected{equation} has a trivial solution] 
$S=\Omega$ with $c_1=0$ (the process never leaves $\Omega$)\corrected{or $S=\emptyset$}. 
%The constant function $1\!\!1_\Omega$ is an eigenfunction of $\infgen$. 
Clearly these trivial solutions are not of any use. How can we solve this conceptual problem? 

When deriving exit rates for biomolecular processes, then the macro states of this system can not be rigorously described as subsets $S$ in micro state space. Rather, transitions are gradual from something that is more inactive to something that can be described as more active. 
We propose replacing \corrected{the indicator function} $1\!\!1_S$ in \eqref{eq:holdprobclassic} by a membership function $\chi:\Omega\rightarrow [0,1]$ which quantifies how much a micro state ${\boldsymbol{x}}\in \Omega$ belongs to the starting macro state.
%is a possible alternative approach. 
The theoretical background of our article, therefore, starts with the \corrected{more general} separation of space and time via 
\begin{equation}
    p_\chi(\boldsymbol{x},t)=\chi(\boldsymbol{x}) \, \mathrm{e}^{-c_1 t}.
\end{equation} 
By definition it satisfies the exponential decay of $p_\chi$ in $t$  and at $t=0$ the probability to be assigned to the starting state is given by $\chi$.
\corrected{Therefore the $\chi$ functions with the smallest exit rates $c_1$ are the most persistent observables or measurements on the system satisfying Markovianity, i.e. time-homogeneous decay. Considering that, however defined, a macro state should be a function of the micro states and exhibit orderly, in our case Markovian, dynamics this}
gives rise to the fundamental idea of identifying the macrostate with the membership function: $\chi$ {\em is} the macro state and the macro state is given in and through $\chi$.
\corrected{Generalising the classical set-based description to that by a $\chi$-function allows us to obtain nontrivial macro-states which have proper exit-rates and therefore allow for a closed description of their dynamics.
In this regard our ansatz can be seen as a necessary consequence of the demand for exact coarse-grained dynamics which is not possible with set-based decompositions.}

At this point we have not provided a method to compute $\chi$ yet. $\chi$ can not be chosen arbitrarily. 
In order to preserve the Markovian long term behavior of the system a projection of the micro system to two different macro states has to be based on an invariant subspace of $\infgen$ \cite{habil, 2008_weber_kube}.
\corrected{Further requiring a decomposition into two complementary (i.e. such that they add up to one) macro states the simplest non-trivial solution is given by  }[which leads
% M. Weber, S. Kube: Preserving the Markov Property of Reduced Reversible Markov Chains. Numerical Analysis and Applied Mathematics, Int. Conf. on Num. Analy. and Appl. Math. 2008, AIP Conference Proceedings, Kos, 1048:593-596, September 2008.
to] 
\begin{equation}\label{eq:Lchi}
\infgen \chi(\boldsymbol{x})= -c_1 \chi(\boldsymbol{x})+c_2(1-\chi(\boldsymbol{x}))
\end{equation}
\corrected{with $c_1, c_2 > 0$ and }[
Here] $\chi$, $1-\chi$ describing the two macrostates.
\corrected{These } can be computed using \corrected{PCCA+ \cite{pcca-plus_grundpaper} or} ISOKANN \cite{isokann_paper, ISOKANNjl} and form an invariant subspace of $\infgen$ which also guarantees long-term consistency with the original dynamics \cite{2008_weber_kube}.
\corrected{Whereas any solution to \eqref{eq:Lchi} leads to a dynamically closed macroscopic description, we will in general aim for solutions with small rates $c_1, c_2$ which represent temporally stable macro states.}

\corrected{In order to derive (\ref{eq:Lchi}) one has to understand how the Markov property of a projected Markov process can be preserved. It means that projection and propagation of the system have to commute. Thus, the projection has to be based on an invariant subspace of $\infgen$.  The constant function is an eigenfunction of $\infgen$. Thus a linear combination of a constant function and of a further eigenfunction leads to a feasible membership function $\chi$. If we apply $\infgen$ to such a linear combination, we arrive at (\ref{eq:Lchi}). For the whole derivation we refer to \cite{ernst}.}
%By solving the last equation, $\chi$ and $1-\chi$ are from an invariant subspace of $\infgen$. ISOKANN \cite{isokann_paper, ISOKANNjl} will be used as a method to generate suitable macro state functions $\chi$.

In Section \ref{chap:grundlagen} of this paper we demonstrate that our macro state formalism based on the $\chi$-function is consistent with the traditional set-based method:
We show that when the $\chi$-functions approach indicator functions the $\chi$ based equations reduce to the classical ones.
%We will show consistency of our $\chi$-function based macro state formalism to the classical set-based approach. We discuss the case when  $\chi$-functions approach indicator functions and show  that the $\chi$ based equations reconstruct the classical ones.
We further motivate its physical meaning by giving a path-based interpretation of the resulting holding probabilities in terms of the Feynman-Kac formula.

In Section \ref{chap:methodik}, we will suggest an approach to apply these theoretic results to extract transition paths from a set of given samples, before demonstrating its application to a molecular system given by the $\mu$-opioid receptor in Section \ref{chap:example}.

%Consistency of our theoretical framework with the set-based approach is given in terms of: The more $\chi$ is an indicator function, the more the presented equations converge to the established ones. Note that $1-\chi(\boldsymbol{x}) \approx 0$ for points which are ``outside'' the starting macro state. 
%However, there is another type of consistency of both equations  (\ref{eq:Lchi}) and (\ref{eq:tmh}) with regard to symmetry. Whatever the equation defining a macro state looks like, it has to be symmetric in the following sense. When exchanging  $S$ by the complement of $S$, we have to replace the exit rate $c_1$ by the exit rate of the complement, which will be denoted as $c_2$. This symmetry also holds for (\ref{eq:Lchi}) when exchanging $\chi$ by the complement $1-\chi$.

\section{Theory of macroscopic exit rates}
\label{chap:grundlagen}

In the following, a consistent theory about macroscopic quantities based on a membership function $\chi:\Omega \rightarrow [0,1]$ is derived. More precisely, the following {\em quantities of interest} are discussed: 
\begin{itemize}
	\item the definition of a macro state via $\chi(\boldsymbol{x})$,
	\item the corresponding position-independent exit rate $c_1$,
	\item the mean holding time $\mht(\boldsymbol{x})$, and 
	\item the reaction direction $\reacdir$ 
 proportional to the gradient $-\nabla \chi$. 
\end{itemize}
Note that computing reaction directions $\boldsymbol{r}$ is possible for sets $S$, too, as described in the introduction. However, for sets we do not get this simple gradient form $-\nabla\chi$. The theory in this section is largely based on \cite{ernst, feynman-kac_paper_weber}. We will extend the theory with statements about consistency and use it to support of the interpretation of $\chi$ as macro states, as explained in the introduction. The computation of $\chi$ will play an important role in our bio-molecular example, see section \ref{chap:example}.

\subsection{Defining macro states via membership functions}
\label{sec:theory}

The starting point in the introduction is that $p_\chi=\chi \mathrm{e}^{-c_1 t}$ would be a way to define holding times of macro states from a conceptual point of view to be able to interpret $c_1$ as an exit rate. The choice of $\chi$ is not arbitrary, but motivated by the necessity of an invariant subspace projection leading to (\ref{eq:Lchi}). At this point it is not yet shown that $p_\chi$ when using the solution $\chi$ of (\ref{eq:Lchi}) is consistent with the stochastic meaning of a holding probability.
We will now demonstrate that it converges to the established definition if $\chi$ becomes the indicator function of a set.

We start by recalling some basic definitions.
Let the state space $\Omega$ of a molecular system comprising $N$ atoms be given as $\Omega=\mathbb{R}^{3N}$
%, i.e. the totality of all states $\boldsymbol{x}_{n \in \{1,2,\dots,N\}} \in \Omega$ that a system can (theoretically) have,
where the position of each individual atom is described by three Cartesian coordinates. 
Let $\rho(\boldsymbol{x},t):\Omega \times \mathbb{R} \to [0, 1]$ denote the probability density distribution of states of the non-linear stochastic dynamics at time $t$. More precisely, the dynamics is given as a Markov process for which an infinitesimal generator 
$\mathcal{L}^\ast: L^1(\Omega)\rightarrow L^1(\Omega)$
(a differential operator, e.g. the Fokker-Planck operator) can be constructed that captures this time-dependent stochastic process. This operator is linear and describes the infinitesimal propagation of $\rho(\boldsymbol{x},t)$:
\begin{align}
	(\mathcal{L}^\ast \rho)(\boldsymbol{x},t) = \frac{\mathrm{d}}{\mathrm{d}t} \rho(\boldsymbol{x},t).
	\label{equ:0}
\end{align}
It also gives rise to its adjoint generator $\infgen: L^\infty(\Omega)\rightarrow L^\infty(\Omega)$ which propagates observables instead of state densities. 
The partial differential equation (\ref{eq:Lchi}) defining $\chi$ is formulated in terms of this adjoint. In this regard $\chi$ can also be interpreted as an observable, i.e. the measurement of the macro state.
To allow for their interpretation as holding probability we are interested in solutions $\chi(\boldsymbol{x}): \Omega \to [0, 1]$ with corresponding  constants $c_1>0, c_2>0$.
\corrected{Assume such $\chi$ and $c_1, c_2$ are given. We then define the $\chi$-holding probability as}[At this point the holding probabilities can be defined by multiplying the solution $\chi$ of \eqref{eq:Lchi} with $\mathrm{e}^{-c_1 t}$,]
\begin{align}
	p_\chi(\boldsymbol{x},t) := \chi(\boldsymbol{x}) \mathrm{e}^{-c_1 t}
	\label{equ:2}
\end{align}
such that \eqref{eq:Lchi} becomes:
\begin{align}
	\mathcal{L}^\ast p_\chi = -c_1 p_\chi + c_2 p_\chi \frac{1-\chi}{\chi}.
	\label{equ:3}
\end{align}
Rearranging results in:
\begin{align}
	\mathcal{L}^\ast p_\chi - c_2 p_\chi \frac{1-\chi}{\chi} = -c_1 p_\chi.
	\label{equ:4}
\end{align}
Due to the relationship
\begin{align}
	\frac{\partial}{\partial t} p_\chi = \frac{\partial}{\partial t} \chi(\boldsymbol{x}) \mathrm{e}^{-c_1 t} = -c_1 \chi(\boldsymbol{x}) \mathrm{e}^{-c_1 t} = -c_1 p_\chi
	\label{equ:5}
\end{align}
Eq. (\ref{equ:4}) can also be represented as follows:
\begin{align}
	\mathcal{L}^\ast p_\chi - c_2 p_\chi \frac{1-\chi}{\chi} = \frac{\partial}{\partial t} p_\chi.
	\label{equ:6}
\end{align}
The solution of the partial differential equation (\ref{equ:6}) together with the initial condition
\begin{align}
	p_\chi(\boldsymbol{x},0) = \chi(\boldsymbol{x}) \mathrm{e}^{-c_1 \cdot 0} = \chi(\boldsymbol{x})
	\label{equ:6b}
\end{align}
can be given in terms of the Feynman-Kac formula \cite{ernst, feynman-kac_paper_weber,feynman-kac_paper_original}:
\begin{align}
	p_\chi(\boldsymbol{x},\tau) = \mathbb{E} \left[ \chi(\boldsymbol{x}_\tau) \cdot \text{exp} \left( -c_2 \int^{\tau}_{0} \frac{1 - \chi(\boldsymbol{x}_t)}{\chi(\boldsymbol{x}_t)} \mathrm{d}t \right) \Big| \boldsymbol{x}_0 = \boldsymbol{x} \right].
	\label{equ:7}
\end{align}
%Here $\mathbb{E}$ is an expectation value. 
\corrected{This representation}[It] allows us to interpret the solution as an expectation over realizations of the stochastic process and therefore builds the bridge from an abstract definition to its interpretation as a probability.

%Let us not focus our interest on the computation of $p_\chi$ itself but on the {\em meaning} of (\ref{equ:7}) for $\chi \rightarrow 1\!\!1_S$. 

\newcommand{\one}{1\!\!1}
To this end let us consider the case where $\chi \approx \one_S$. \corrected{The expectation is to be taken over all trajectories starting in $\boldsymbol{x}_0$.}[Then $p_\chi$ is the expectation over trajectories starting in $\boldsymbol{x}_0$.]
%\corrected{to stay in $S$ until time $\tau$ at least}. 
Once any such trajectory leaves the set $S$ the integral becomes infinite and exponential function evaluates to $0$. Otherwise the exponential stays $1$, as well as $\chi(\statex_\tau)=1$.
We therefore recover the definition of classical holding probability $p_\chi(\statex, \tau) = p_S(\statex, \tau)$ in \eqref{eq:holdprobclassic}\corrected{, i.e. the probability to stay in $S$ for time $\tau$ at least}, see also (9) and (10) in \cite{ernst}, as well as (3.31) in \cite{schuette}.

\corrected{We summarize this result in the following proposition:
\begin{proposition}Let $\chi \approx \one_S$ be a solution to \eqref{eq:Lchi}. Then the $\chi$-holding probability approximates the classical holding probability of $S$:
$$p_\chi(\statex, t) \approx p_S(\statex, t).$$
\end{proposition}
}

%\textcolor{gray}{
%Think of an evaluation of $p_\chi(\boldsymbol{x},\tau)$ which uses the simulation of several trajectories starting in $\boldsymbol{x}$ within a time interval $[0,\tau]$ according to the stochastic dynamics. Let us  understand how (\ref{equ:7}) is to be interpreted if $\chi$ is close to an indicator function. Assume $M_0$ and $M_1$ are pathwise connected sets forming a partition of $\Omega$:
%\begin{align}
%	M_0 \, \cup \, M_1 = \Omega , \; M_0 \, \cap \, M_1 = \emptyset.
%	\label{equ:7b}
%\end{align}
%We are looking for a solution of the partial differential equation (\ref{eq:Lchi}) where $\chi$ is as crisp as possible and close to an indicator function of $M_1$. The function value of the membership function $\chi(\boldsymbol{x})$ approximately quantifies the degree of membership $\in [0;1]$ of the point with regard to the subset $M_1$ \cite{goguen1973}.
%\begin{align}
%	\chi(\boldsymbol{x}) \approx 0 \; \forall \, \boldsymbol{x} \in M_0 , \; \chi(\boldsymbol{x}) \approx 1 \; \forall \, \boldsymbol{x} \in M_1.
	\label{equ:7c}
%\end{align}
%For a simulation of the dynamical system with stopping  time $\tau$, that means a trajectory in $\Omega$, three cases can occur:
%\begin{enumerate}
%	\item The trajectory ends at a point $\boldsymbol{x}_\tau \in M_1$ and all points of the trajectory lie in $M_1$. We then have $1 - \chi(\boldsymbol{x}_t) \approx 0$ in (\ref{equ:7}) over the entire period and consequently the exponential function $\approx 1$ and $\chi(\boldsymbol{x}_\tau) \approx 1$. Then this trajectory is counted for $\mathbb{E}$.
%	\item The trajectory ends at a point $\boldsymbol{x}_\tau \in M_1$ and some of the points of the trajectory lie in $M_0$. Then also $\chi(\boldsymbol{x}_\tau)\approx 1$, but for part of the time the denominator in the integral is $\chi(\boldsymbol{x}_t) \approx 0$, so the value of the integral approaches $\infty$ and therefore the exponential function is close to $0$. This trajectory is not counted for $\mathbb{E}$.
%	\item The trajectory ends at a point $\boldsymbol{x}_\tau \in M_0$. (This means that some of the points of the curve must also lie in $M_0$.) Then the exponential function is again $\approx 0$ and also $\chi(\boldsymbol{x}_\tau) \approx 0$. Both alone or together lead to not counting this trajectory for $\mathbb{E}$.
%\end{enumerate}
%In equation (\ref{equ:7}) only those trajectories are counted for which all points lie in $M_1$ over the whole time interval $[0,\tau]$. These are counted in relation to the number of all started simulations. This means that the quantity $p_\chi(\boldsymbol{x},t)$ is close to the holding probability of $M_1$. 
%}

With regard to this interpretation, $p_\chi(\boldsymbol{x},t)$ is seen as the holding probability of the macro state $\chi$. Due to the separated term $\mathrm{e}^{-c_1 t}$ in (\ref{equ:2}), the holding probability decreases exponentially with the decay constant $c_1$. This means that $c_1$ is the exit rate from $\chi$. Since the function value of $p_\chi(\boldsymbol{x},0)=\chi(\boldsymbol{x})$ is interpreted as a holding probability, it is necessary that $\chi$ can only take values in the interval $[0,1]$.

The time integral over the holding probability is the mean holding time $\mht(\boldsymbol{x})$, which is proportional to the inverse of the exit rate $c_1$:
\begin{align}
	\mht(\boldsymbol{x}) = \int^{\infty}_{0} p_\chi(\boldsymbol{x},t) \mathrm{d}t = \int^{\infty}_{0} \chi(\boldsymbol{x}) \mathrm{e}^{-c_1 t} \mathrm{d}t = \lim_{t \to \infty} \left[ -\frac{1}{c_1} \chi(\boldsymbol{x}) \mathrm{e}^{-c_1 t} \right]^{t}_0 = \frac{1}{c_1} \chi(\boldsymbol{x}).
	\label{equ:8}
\end{align}

%%%%%%%%%%%%%%%%%%%%%%%%
% %Idea for a new version:
% \textcolor{gray}{
% In addition, the reaction path can be derived from $\mht(\boldsymbol{x})$. If $\mht$ is the expected time till ``a reaction takes place'', then the vector field $\boldsymbol{r}(\boldsymbol{x})$ points into the direction in which this time decreases the most. Reaction paths (in this special case, the atomic positions over the course of the reaction) are the curves obtained integrating $\boldsymbol{r}(\boldsymbol{x})$, which is proportional to the gradient of $\chi$:
% \begin{align}
% 	\boldsymbol{r}(\boldsymbol{x}) = -\nabla \mht(\boldsymbol{x}) = -\frac{1}{c_1} \nabla \chi(\boldsymbol{x})  \propto -\nabla \chi(\boldsymbol{x}).
% 	\label{equ:9}
% \end{align}
% $\boldsymbol{r}$ has this simple form because of the time separation of $p_\chi$, which was our theoretical starting point. By our interpretations of membership, $\chi$ is now like the reaction coordinate of our system. It therefore can be seen as an ordering of micro states by their mean holding times, which will be exploited in section \ref{chap:methodik} below.
% }
%%%%%%%%%%%%%%%%%%%%%%%%

The mean holding time immediately leads to a definition of a reaction direction:
Following the gradient of $\mht$ increases the time until ``a reaction takes place''.
Therefore by defining the reaction direction $\reacdir:\Omega\rightarrow\mathbb{R}^{3N}$,
\begin{align}
	\boldsymbol{r}(\boldsymbol{x}) = -\nabla \mht(\boldsymbol{x}) = -\frac{1}{c_1} \nabla \chi(\boldsymbol{x})  \propto -\nabla \chi(\boldsymbol{x}),
	\label{equ:9}
\end{align}
we obtain a vector field along which the mean holding time decreases uniformly and which is proportional to $\nabla \chi$.
This also means that $\chi$ itself can be understood as an order parameter, i.e. a reaction coordinate for the system.
Note that we obtain this time independent result only as a consequence of the initial time separation ansatz for $p_\chi$.
By integrating curves tangential to $\reacdir$ one can obtain reaction paths in $\Omega$.
In Section \ref{chap:methodik} we will make use of the order parameter interpretation to subsample a representative reactive path from a given pool of simulation data.

% In addition, the reaction direction can be derived from $\mht(\boldsymbol{x})$. 
% If $\mht$ is the expected time until ``a reaction takes place'', then the reaction direction $\boldsymbol{r}(\boldsymbol{x})$ should point into the direction in which this time decreases the most. As such it is proportional to the gradient of $\chi$:

% Note here that $\boldsymbol{r}$ admits this simple form because of the time separation of $p_\chi$, which was our theoretical starting point. By our interpretations of membership, $\chi$ is now like the reaction coordinate of our system. It therefore can be seen as an ordering of micro states by their mean holding times, which will be exploited in section \ref{chap:methodik} below.

The possibility of calculating these quantities of interest from (\ref{eq:Lchi}) is a motivation to develop an efficient method for solving this equation. We will now show how to express these quantities in terms of the Koopman operator, before discussing ISOKANN, an algorithm for their computation.

\subsection{Membership functions from Koopman operator}

So far the description of $\chi$ was based on the infinitesimal generator $\infgen$ but a suitable analytical solution of the corresponding partial differential equation \eqref{eq:Lchi} is not available. However, it is possible to transform (\ref{eq:Lchi}) into an equation for which a constructive solution is possible. We will now show how we can similarly formulate it in terms of the Koopman operator $\koop$ and how we can switch between the formalisms \corrected{and work out the relation between $\chi$ function and eigenfunctions of $\koopt$}. The problem of actually computing $\chi$ will then be addressed in the next subsection.

The Koopman operator $\mathcal{K}^\tau$ is the time-integral or solution operator of $\mathcal{L}^\ast$ for some lag-time $\tau > 0$ and can be formally defined as $\mathcal{K}^\tau = \mathrm{e}^{\tau \mathcal{L}^\ast}$.
%The algorithmic solution of this partial differential equation will make use of the Koopman operator $\mathcal{K}^\tau$ which is the integral of the infinitesimal generator $\mathcal{L}^\ast$:
% \begin{align}
% 	\mathcal{K}^\tau = \mathrm{e}^{\tau \mathcal{L}^\ast},
% 	\label{equ:10}
% \end{align}
It can also be defined by its action on observable functions $f:\Omega \rightarrow\mathbb{R}$:
\begin{align}
	(\mathcal{K}^\tau f)(\boldsymbol{x}) := \mathbb{E}\big[f(x(\tau))\mid x(0)=\boldsymbol{x}\big],
\end{align}
where the expectation is taken over independent realizations $x$ of the process, starting in $x(0) = \boldsymbol{x}$.
%where $\boldsymbol{x}(0)$ represents the starting point and $\boldsymbol{x}(\tau)$ represents end points of independent realizations of the process.
\corrected{It can be understood as the} expected measurement $\koop^\tau f$ of an observable $f$ after a lag time $\tau$.
\corrected{Being an expectation value}[In this way,] its action can be approximated using Monte-Carlo estimates over simulations, making it particularly suitable for applications.

To transform (\ref{eq:Lchi}) into an equation for $\koop^\tau$ we substitute
\begin{align}
	\alpha = c_1 + c_2
	\label{equ:11}
\end{align}
to arrive at the following equation:
\begin{align}
	\mathcal{L}^\ast \chi = -\alpha \chi + c_2.
	\label{equ:12}
\end{align}
This shows that $\infgen$ acts as a shift-scale operator\corrected{on $\chi$} if and only if $\chi$ solves \eqref{eq:Lchi}. 
%Now we make use of the Koopman operator: 
%\begin{align}
%	\mathcal{K}^\tau \chi = \mathrm{e}^{\tau \mathcal{L}^\ast} \chi.
%	\label{equ:13}
%\end{align}
Making use of the formal exponential representation of $\koop^\tau$ 
%into the following equation using (\ref{equ:12}) 
and its series expansion one obtains \cite{ernst, feynman-kac_paper_weber}:
\begin{align}
	\mathcal{K}^\tau \chi = \mathrm{e}^{-\tau \alpha} \chi + \frac{c_2}{\alpha} (1-\mathrm{e}^{-\tau \alpha}).
	\label{equ:14}
\end{align}
Setting 
\begin{align}
	\gamma_1 = \mathrm{e}^{-\tau \alpha}, \quad \gamma_2 = \frac{c_2}{\alpha} (1-\gamma_1)
	\label{equ:gamma}
\end{align}
this becomes:
\begin{align}
	\mathcal{K}^\tau \chi = \gamma_1 \chi + \gamma_2.
	\label{eq:isosolution}
\end{align}
So, just as with equation (\ref{equ:12}), $\koop^\tau$ acts as a shift-scale if and only if $\chi$ is a solution to \eqref{eq:Lchi}.
\corrected{
Noting that $\koopt \one = \one$ we further see that $f:=\chi - \frac{c_2}{\alpha}$ is an eigenfunction of $\koopt$ with eigenvalue $\gamma_1$:
\begin{align}
    \koopt \left(\chi - \frac{c_2}{\alpha}\right) = \gamma_1 \chi + \frac{c_2}{\alpha}(1-\gamma_1) - \frac{c_2}{\alpha} = \gamma_1\left(\chi - \frac{c_2}{\alpha}\right).
\end{align}
These findings are summarized in the following proposition:
\begin{proposition}
    Let the parameters $c_1, c_2, \alpha, \gamma_1, \gamma_2$ satisfy their relations above. The following are equivalent:
    \begin{itemize}
        \item $\chi$ solves the ISOKANN problem \eqref{eq:Lchi}.
        %\begin{equation*}
        %    \infgen \chi(\boldsymbol{x})= -c_1 \chi(\boldsymbol{x})+c_2(1-\chi(\boldsymbol{x})).\end{equation*} 
        \item $\infgen$ acts as a shift-scale on $\chi$
        %\begin{equation*}\mathcal{L}^\ast \chi = -\alpha \chi + c_2\end{equation*}
        with scale $-\alpha$ and shift $c_2$.%, $\alpha = c_1 + c_2$.
        \item $\koopt$ acts as a shift-scale on $\chi$
        %\begin{equation*}
            %\mathcal{K}^\tau \chi = \gamma_1 \chi + \gamma_2
        %\end{equation*}
        with scale $\gamma_1$ and shift $\gamma_2$. %$\gamma_1 = \mathrm{e}^{-\tau \alpha}, \quad  = \frac{c_2}{\alpha} (1-\gamma_1)$.
        \item $\chi - \frac{c_2}{\alpha}$ is an eigenfunction of $\koopt$ with eigenvalue $\gamma_1$.
    \end{itemize}
\end{proposition}
%To summarize, $\chi$ solving \eqref{eq:Lchi} and $\infgen$ resp. $\koopt$ acting as a shift-scale on $\chi$ are all equivalent, where the latter have the scale and shift parameters $\alpha$ and $c_2$ resp. $\gamma_1$ and $\gamma_2$.
}[Note however, that its shift  ($\gamma_1$) and scale ($\gamma_2$) parameters differ.]

%We saw that equation shows that $\chi$ is a solution iff $\mathcal{L}^\ast \chi$ is the same as scaling $\chi$ (with $-\alpha$) and then shifting it (with $c_2$).
%Equation (\ref{eq:isosolution}) shows that the same holds for its exponential $\mathcal{K}^\tau$ which scales with $\gamma_1$ and shifts with $\gamma_2$.
The above identities allow us to switch between the infinitesimal generator and Koopman framework. In particular, we can compute the exit rate $c_1$ from the Koopman parameters $\gamma_1$ and $\gamma_2$:
%The ISOKANN method \cite{isokann_paper, sikribweb, ISOKANNjl} is a fixed-point iteration in which iterative scaling and shifting takes place so that the function values are constraint to the interval $[0,1]$ for all samples. If ISOKANN converges, a $\chi$ is found that solves (\ref{eq:Lchi}), (\ref{equ:12}) and (\ref{eq:isosolution}). 
%In addition, $\gamma_1$ and $\gamma_2$ are also found in the convergence case,
%Rearranging (\ref{equ:gamma}) gives
\begin{align}
	\alpha = -\frac{\text{ln}\gamma_1}{\tau}, 
 \quad c_2 = \frac{\alpha\gamma_2 }{1-\gamma_1}, 
 \quad c_1 = \alpha - c_2.
	\label{equ:16}
\end{align}
% which together with (\ref{equ:11}) results in:
% \begin{align}
% 	c_1 = \alpha - c_2.
% 	\label{equ:18}
% \end{align}
This allows to estimate the \corrected{respective constants, e.g. the exit rate $c_1$, }[exit rate]
by evaluating $\chi$ and $\koopt\chi$ at sample points $\statex\in\Omega$ 
%we can therefore estimate the exit rate $c_1$ by
and solving the linear regression problem \eqref{eq:isosolution}.
%This means that all the relationships required to calculate the quantities of interest are known.

\subsection{ISOKANN for computing membership functions}

The ISOKANN (Invariant subspaces of Koopman operators learned by a neural network) method \cite{isokann_paper, sikribweb, ISOKANNjl} is a fixed-point iteration which combines the use of a neural network for representing the high-dimensional $\chi$ function with the Koopman formalism to enable its training on simulation data.

%in which iterative scaling and shifting takes place so that the function values are constraint to the interval $[0,1]$ for all samples. 
On convergence it returns $\chi$ that solves (\ref{eq:Lchi}), (\ref{equ:12}) and (\ref{eq:isosolution}). 
%We briefly explain, how ISOKANN works. 
Solving the partial differential equations involving $\mathcal{L}^\ast$ directly is not feasible due to the high dimensionality of molecular systems.
Molecular simulations on the other hand enable us to estimate the action of $\mathcal{K}^\tau$ on an observable.
%What is needed to solve these equations?
%$\mathcal{L}^\ast$ is a differential operator. Trying to solve a high-dimensional partial differential equation leads to numerical problems. Whereas, using molecular simulations enables for the computation of the action of $\mathcal{K}^\tau$ to a function $\chi$ point-wise. 
For this reason, ISOKANN attempts to solve (\ref{eq:isosolution}) by using the action of $\mathcal{K}^\tau$ for the calculation of $(\mathcal{K}^\tau \chi_i)(\boldsymbol{x})$, where $\chi_i$ is the $i$-th iterate of $\chi$ and $\boldsymbol{x}$ is a training point. It approximates the expectation value
\begin{align}
	(\mathcal{K}^\tau \chi_i)(\boldsymbol{x}) := \mathbb{E}\big[\chi_i(x(\tau))\mid x(0)=\boldsymbol{x}\big],
	\label{equ:koopman}
\end{align}
by a Monte-Carlo estimate over trajectory simulations $x$ starting in different starting points $\statex$.
%where the starting function $\chi_0:\Omega \to \mathbb{R}$ is chosen at random.
The next iterate is then given by the shift-scaled $\koopt \chi_i$:
%Reversing the scale shift procedure on the right hand side of (\ref{eq:isosolution}) provides the next iterate:
\begin{align}
	\chi_{i+1} := \frac{\phantom{\Vert}\mathcal{K}^\tau \chi_i - \min(\mathcal{K}^\tau \chi_i)\phantom{\Vert_\infty}}{{\Vert \mathcal{K}^\tau \chi_i - \min(\mathcal{K}^\tau \chi_i)\Vert}_{\infty}},
	\label{equ:power}
\end{align}
which is motivated by inverting the shift and scale of \eqref{eq:isosolution}, such that the solution $\chi$ to \eqref{eq:isosolution} is indeed a fixed point. \corrected{The initial guess $\chi_0$ is chosen randomly.}
ISOKANN is based on the power method, an iterative method used to obtain the dominant eigenfunction of a linear operator. In ISOKANN additional scaling and shifting in each iteration ensures that it does not converge to the constant function, but against the membership function $\chi(\boldsymbol{x}):\Omega \to [0, 1]$ \cite{sikribweb} -- similar to targeting the second eigenvalue in the inverse power method.
%In the algorithmic scheme, minima and maxima in (\ref{equ:power}) are selected according to a set of test points used for the training of $\chi$.
%Each iterate $\chi_i$ is approximated by an artificial neural network (NN) which is trained on a supervised manner on samples $x$.
In order to represent the iterates $\chi_i$, we approximate them by a neural network.
The equality assignment in the iteration \eqref{equ:power} thus becomes a supervised learning problem at data points $\statex$ with labels given by the corresponding right hand side of \eqref{equ:power} evaluated on the previous generation of the network.
The iterations are then terminated by a stopping criterion which
%which is adapted by training in each iteration step (that means that the previous state of the ANN continues to be used) until a stopping criterion is reached.
%This 
can be either a high correlation coefficient $(\chi_i(\boldsymbol{x}_n),\chi_{i+1}(\boldsymbol{x}_n))_n$ or a small empirical loss $\|\chi_i - \chi_{i+1}\|_2$, indicating that we found an approximate solution to \eqref{eq:isosolution}.

Assuming infinite data and perfect representation by the neural network, this iteration indeed converges to the \corrected{}[slowest] solution $\chi$ of \eqref{eq:isosolution}\cite{sikribweb}
\corrected{ with the smallest rate $c_1$. It is spanned by the constant eigenfunction with eigenvalue 0 and the second eigenfunction with eigenvalue closest to 0.
%and therefore resulting in the slowest possible mixing with its complement $1-\chi$.
%This means that $\chi_i$ converges towards a linear combination of the constant function (the eigenfunction of $\infgen$ for eigenvalue 0) and the second eigenfunction with eigenvalue closest to $0$.
This solution is unique, if the corresponding eigenvalue is simple.} In practice however, for \corrected{a cluster of}[multiple] low-lying eigenvalues, this routine can result close to one of multiple possible membership functions (each representing one of these slower processes). In that case the procedure can be repeated and the resulting membership functions can be used to construct an invariant subspace of $\koop$.

%It stops, if the linear correlation coefficient of the pairs $(\chi_i(\boldsymbol{x}_n),\chi_{i+1}(\boldsymbol{x}_n))_n$ reaches a sufficiently large value,  such that the action of ${\cal K}^\tau$ corresponds to a shift-scale operation. If high correlation is reached, the equation (\ref{eq:isosolution}) can be regarded as valid. 
%The whole procedure can be repeated several times with different starting functions, such that several membership functions $\chi_{end}$ are available as solutions, which are usually linearly independent. All found solution functions together with the constant function span  a $\mathcal{K}^\tau$-invariant subspace in function space.
%, because the right hand side of (\ref{eq:isosolution}) is a linear combination of such a function and a constant function. 
One of ISOKANN's main benefits is that it avoids discretizing the state space, which is crucial for the application to high-dimensional systems and avoiding the curse of dimensionality
\cite{isokann_paper}. In the ISOKANN algorithm it is possible to train artificial neural networks by collecting many short-term trajectories of simulation length $\tau$ from different starting points in high-dimensional spaces. 
Thus, ISOKANN can be applied to many independent short-time trajectories or it can even decide where in $\Omega$ to enrich simulation data \cite{sikribweb}. 
In our illustrative example we will apply it to a small number of medium-length trajectories. The resulting $\chi$ function will then be used to subsample a reactive path from the data, which leads us to the next section.

\section{Extracting transition paths from simulations}
\label{chap:methodik}
\corrected{Once we obtained a $\chi$ function from ISOKANN we might be tempted to directly compute a reaction path following the gradient of $\chi$ as in \eqref{equ:9}.
This however is problematic as the neural network approximation is good only in the region of sufficient training data. The gradient of $\chi$ however could point away from this region, accumulating more and more approximation errors and quickly lead to unobserved unphysical states.  One might solve this problem by projecting back to the physical regime, e.g. via energy minimization. This however requires access to the original systems potential/forces and is not only computationally expensive but also involved from an implementation perspective.}

\corrected{We now propose an alternative method which can be applied as a post-processing step to already sampled simulation data without requiring further information about the system. 
Applied as a post-processing step to molecular
simulations, together with ISOKANN, it provides researchers with a direct step-by-step representation of the slowest process (which is contained in the simulation
data) by identifying this process and filtering out the intermediate fluctuations.}

To this end we understand the learned $\chi$ values as an order parameter for a set of simulation data in $\Omega$. 
%In this section we show how to apply the $\chi$ function trained by ISOKANN. 
As shown in (\ref{equ:8}), the membership values are interpreted to be proportional to mean holding times of a macro state. 
%The $\chi$ values represent a ``temporal'' ordering of micro states. 
The level sets of the function $\chi(\boldsymbol{x})$ in this regard  correspond to micro states $\boldsymbol{x}$ which ``take place simultaneously'' in this newly defined time axis. 
By interpreting this mean-holding-time as a ``temporal'' parameter we can extract macroscopic transition paths from simulation data. \corrected{Replacing the time from the simulation by $\chi$ furthermore allows us to incorporate data of different simulations, treating them merely as $\chi$ ordered data points and thus allowing for higher resolution of the results.}
%In our illustrative example in section \ref{chap:example}, we aim to find \corrected{$\chi$-ordered}[temporally ordered] paths in $\Omega$. Thus, micro states will be picked forming this path out of given molecular simulation data. 
%In this context, this approach can be seen as a post processing step to classical molecular simulation by which a representation of the slow dynamics contained in the data is obtained. 

\corrected{The core problem of choosing a path through the data samples consists of balancing the progress in the reaction direction, $\nabla\chi$, versus spatial movement on the levelsets of $\chi$. More formally we look for an }[
The $\chi$ values will be interpreted as order parameter or artificial time of the transition between macro states.
An] ordered list $I \subset J=\{1,\dots, N\}$ of samples from simulation points $X_J = (\boldsymbol{x}_j)_{j \in J}$ in $\Omega$ \corrected{}[is searched for], such that $\chi(\boldsymbol{x}_j)$ is an increasing sequence with smooth spatial transitions, i.e. without large deviations \corrected{$\statex_{j+1} - \statex_j$}[of $\Delta \boldsymbol{x}_j$].
\corrected{To this end we will model the spatial movement as a Brownian motion through the time-parameter $\chi$.}
[To this end the trajectory samples $\boldsymbol{x}_j$ are modelled locally as the result of a Brownian motion $dX_t = \sigma dW_t$ with their respective $\chi$-value as time.
Note here that we $t$ does not denote the simulated time but the progression in the reaction direction, i.e. $t=\chi$.]

\corrected{For the classical Brownian motion $dX_t = \sigma dW_t$}
the probability of obtaining a specific set of samples,\corrected{ conditioned on the sampling time points $t_i$,} is given by the finite dimensional distribution \cite{pavliotis} (in our case we assume $\boldsymbol{b}\equiv 0$ and $\sigma(\boldsymbol{x}) \equiv \sigma$ as well as $t=\chi$):
% \begin{align}
%     p(x_1,\ldots,x_n) = \left( \prod^{n-1}_{i=1} \frac{1}{\sqrt{2\pi\sigma(x_i)^2\Delta t_i}} \right) \exp\left(-\sum^{n-1}_{i=1} L\left(x_i,\frac{x_{i+1}-x_i}{\Delta t_i}\right) \, \Delta t_i \right)
% \end{align} 

\begin{align}
    p(\boldsymbol{x}_1,\ldots,\boldsymbol{x}_n | t_1, \dots, t_n) = 
    \left( \prod^{n-1}_{i=1} (2\pi\sigma^2\Delta t_i)^{-d/2} \right)
    \exp\left(-\sum^{n-1}_{i=1} \frac{||\boldsymbol{x}_{i+1} - \boldsymbol{x}_i||^2}{2\sigma^2 \Delta t_i} \right).
    \label{eq:likelihood}
\end{align}

\begin{figure}[t]
    \centering
    \hfill
    \begin{minipage}[b]{0.45\textwidth}
        \centering
        \includegraphics[width=\textwidth]{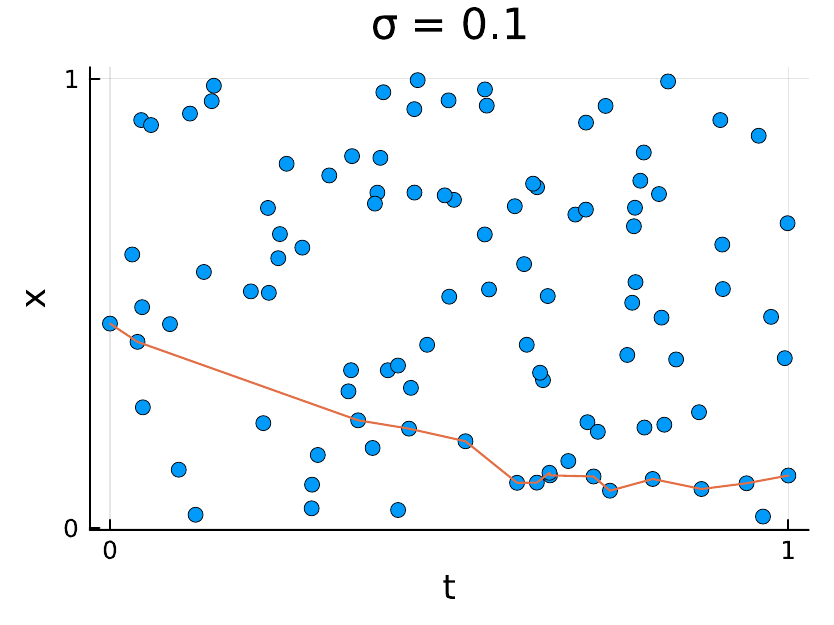}
        
    \end{minipage}
    \hfill
    \begin{minipage}[b]{0.45\textwidth}
        \centering
        \includegraphics[width=\textwidth]{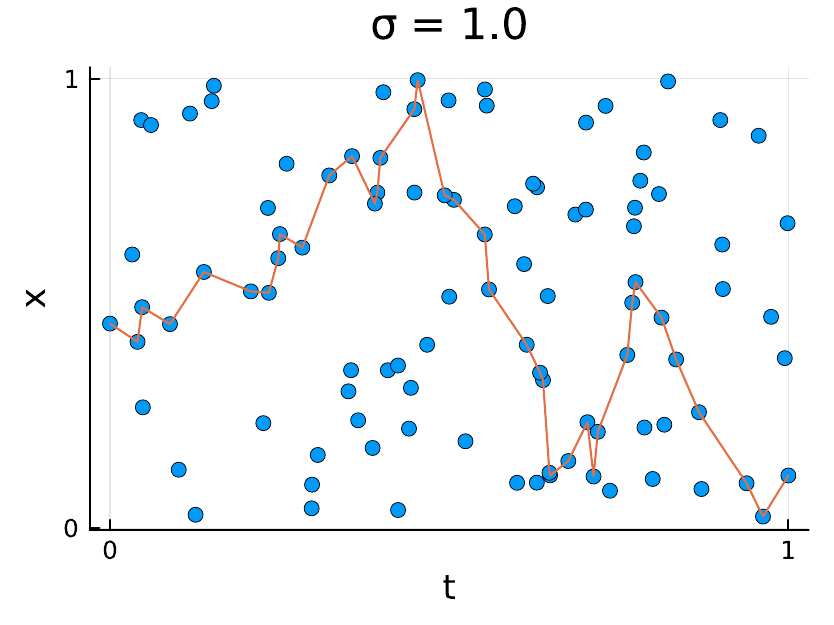}
        
    \end{minipage}
    \hfill
    \caption{\corrected{Illustration of the maximum likelihood path \eqref{eq:likelihood}}  on 100 points distributed uniformly in space and time. For lower $\sigma$ values (left) the path prefers less jumps with small spatial displacement, whilst for higher $\sigma$ (right) the path goes through more points at the cost of more erratic movement. \corrected{Note that this is just an illustration of the maximum-likelihood-path with synthetic data in classical time $t$ (which in our application will be replaced by $\chi$).}}
    \label{fig:shortestpath}
\end{figure}

This formula is typically used to obtain the probability given a specific set of $\Delta t_i=t_{i+1} - t_i$.
\corrected{In our case we will use it to also compare different sampling times $t_i$ by setting $p(\boldsymbol{x}_1,\ldots,\boldsymbol{x}_n | t_1, \dots, t_n) = p(\boldsymbol{x}_1,\ldots,\boldsymbol{x}_n, t_1, \dots, t_n)$ which allows to balance temporal with spatial jumps.}
\corrected{This can be justified}[
We expect that can be understood] from a Bayesian perspective as prescribing a uniform prior on the number and length of time intervals\corrected{(as we have no preference for specific $\Delta t$ values to appear in the solution) although}[
For the time being, we consider it a heuristic justified by the results, but] further investigation \corrected{of this view} may be warranted.
The parameter $\sigma$ plays the role of a smoothing parameter, balancing the likelihood of jumps in space or time.
A high $\sigma$ allows for more erratic jumps over short time-spans while a lower $\sigma$ favors spatially closer jumps possibly necessitating longer time-spans, as illustrated in Figure \ref{fig:shortestpath}.
%with $L(x,v) = \frac{1}{2} ((v-b(x))/\sigma(x))^2$ for a system $dX_t = b(X_t)dt + \sigma(X_t) dW_t$ \cite{WIKIPEDIA?}.

\corrected{We now continue to find the maximum likelihood path through our sampling data by replacing the classical time parameter with the samples $\chi$ value, such that $\Delta t_i = \chi(x_{i+1}) - \chi(x_i)$.}

Taking the logarithm allows us to transform \eqref{eq:likelihood} into a sum of log probabilities, $\log p(\boldsymbol{x}_1,\dots, \boldsymbol{x}_n) = \sum_{i=1}^n \log p(\boldsymbol{x}_i, \boldsymbol{x}_{i+1})$ with
\begin{align}
    \log p(\boldsymbol{x}, \boldsymbol{y}) 
    = \log \left( \left(2\pi\sigma^2\left(\chi(\boldsymbol{y})-\chi(\boldsymbol{x})\right)\right)^{-d/2} \right)
    %\log \left( \frac{1}{\sqrt{2\pi\sigma^2 (\chi(y)-\chi(x))}} \right)
    - \frac{||\boldsymbol{y}-\boldsymbol{x}||^2}{2\sigma^2(\chi(\boldsymbol{y})-\chi(\boldsymbol{x}))}.
\end{align}

Finding the maximal likelihood path then corresponds to solving the shortest path problem from a (set of) point(s) $\chi(\boldsymbol{x}_i) \approx 0$ to a (set of) point(s) $\chi(\boldsymbol{x}_i) \approx 1$ with edge distance $e_{ij}$ between two points (nodes)
\begin{align}
    e_{ij}=\begin{cases}
    -\log p(\boldsymbol{x}_i, \boldsymbol{x}_j) & \text{if } \chi(\boldsymbol{x}_i) < \chi(\boldsymbol{x}_j) \\
    \infty & \text{otherwise},
\end{cases}
\end{align}
where transitions forward in time (i.e. increasing $\chi$-value) are enforced by the corresponding $\infty$ weight.
The shortest path problem can be solved with the Bellman-Ford algorithm \cite{bellmanford}.

Of course the assumption of uniform Brownian motion for the underlying dynamics is far-fetched for an actual molecular system.
However, with only finite simulation data, the locally possible jumps will be dictated mainly by the available data \corrected{which already incorporates the physical drift}. The Brownian assumption introduces only a small bias which is largely negligible for the choice of paths in the temporal, i.e. $\chi$ direction, as any larger deviations from the Brownian distribution will already be reflected in the available data.
\corrected{
We believe this can be improved in cases where the acceleration (force and masses) and the diffusion coefficient $\sigma$ of the dynamics are known. This however is not straightforward, as we have replaced the ordinary time with $\chi$ which requires some projection of the stochastic process.}

In this regard our proposed algorithm can be understood as a simple heuristic filtering method to obtain a smooth path through already provided samples along the learned reaction coordinate from $\chi=0$ to $\chi=1$ (or vice versa).
\corrected{The result is the most-likely path (under the Brownian assumption) between the extremal conformations identified by $\chi$. Being composed of the actually simulated, hence physically relevant, data it provides a smooth transition through the identified slowest process indexed by the mean holding time of the corresponding macro-state.}

%\corrected{The extracted path is a physically possible scenario. However, the question remains as to the probability of such a path occurring. If you only look at a local area of the path, then the next steps are not improbable. But if you look at the entirety of the path, then it is an improbable result of a simulation because there are no detours. However, a researcher who wants to understand the course of a (biological) process is actually interested in seeing the path without detours, because they distract from the essentials, from the relevant steps of a process. And this is precisely our motivation for calculating the path: To present the process in a direct and straightforward way.}

\section{Illustrative Example: Opioid Receptor} 
\label{chap:example}

To illustrate the added value of $\chi$ computation, we will show an molecular dynamics (MD) example which is part of a pharmaceutical project. We first learned the $\chi$ function from molecular simulation trajectories and applied the described reaction path extraction along $\chi$ to a high-dimensional molecular system consisting of 4734 atoms. The application background is given by understanding pain relief using opioids. Strong painkillers like morphine and fentanyl act upon a special type of receptor in the body known as the $\mu$-opioid receptor (MOR). This receptor is part of the family of opioid receptors and is a G-protein coupled receptor found in various parts of the body such as brain, spinal chord, and gastrointestinal tract \cite{pert1973opiate}. The indiscriminate activation of the MOR across the whole body is one of the causes of severe side-effects of this family of strong pain killers. \cite{darcq2018opioid}.

However, it is proven that chemical changes at site of inflammation cause the creation of a micro-environment \cite{reeh1996tissue}. The knowledge about the local micro-environment can be used to design peripherally restricted strong pain killers with potentially less side-effects \cite{nontoxic_pain_killer}. Different micro-environments may lead to different dynamics of the MOR. One possible chemical change of the MOR in inflamed tissue postulates the formation of disulfide bonds as the concentration of reactive oxygens species goes up.

\subsection{Algorithmic details}

\begin{figure}[t]
    \centering
    \includegraphics[width=0.8\linewidth]{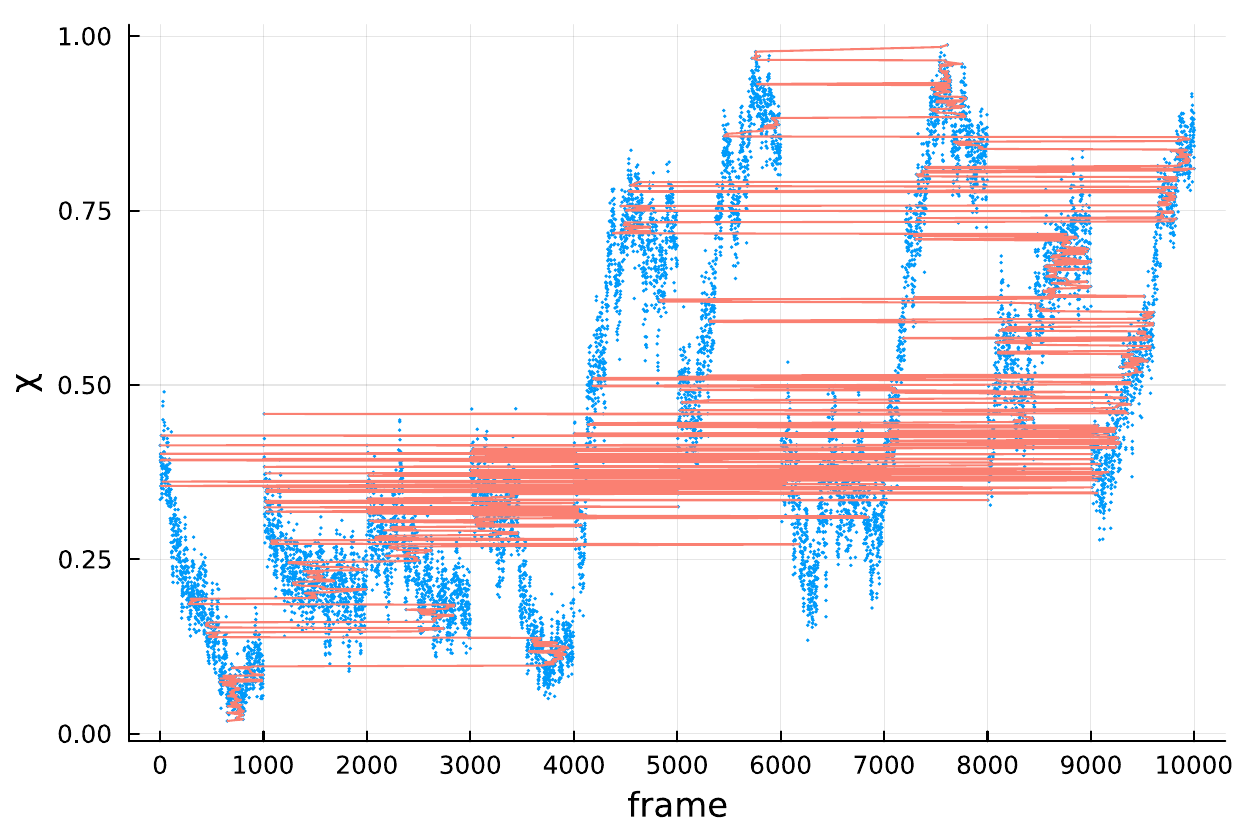}
    \caption{
    %This plot shows (blue dots) the values of $\chi(\statex)$ for the micro states $\statex$ of the 10 independent simulations (each simulation has 1000 ``frames''). One can see that each single simulation takes samples from different parts of the $\chi$-interval $[0,1]$. The orange line shows the extracted reaction path along ordered $\chi$-values. However, due to small spatial distance in $\Omega$ sometimes the orange line ``jumps'' between the 10 trajectories.
    The membership value $\chi(x)$ for each state (frame) $x$ obtained from 10 independent simulations, with each simulation comprising 1,000 frames.
    The y-axis represents the macroscopic transition, showing that different trajectories cover distinct segments of this transition while exhibiting partial overlap with other trajectories.
    The extracted reaction path (orange line) progresses \corrected{monotonously} from $\chi \approx 0$ to 
    $\chi \approx 1$ while maintaining \corrected{(not depicted)} spatially smooth transitions .
    Note that it incorporates the data from amongst all 10 simulations and even jumps \corrected{between} them where facilitated by small spatial distances in $\Omega$.
    }
    \label{fig:chi_along_time}
\end{figure}

%Let us outline our whole pipeline:
Our input data is taken from $10$ different simulations of \corrected{the} MOR. After the primary proximity analysis on the active structure of the MOR (\corrected{Protein Data Bank (PDB) ID}[PDB] 8EF5) a disulfide bond was introduced in the inactive structure of the MOR in between location CYS159 and CYS251 (\corrected{Protein Data Bank ID}[PDB] 7UL4) \cite{zhuang2022molecular,robertson2022structure}. %(Human MOR equivalent 161 and 253) 
10 simulations (with explicit water and with a lipid-bilayer for the MOR) were run. Each simulation spans an interval of 100 nanoseconds, totalling to 1 microsecond. 

After simulation, pairwise distances over all $\alpha-$carbons that can be observed to be closer then a threshold ($d_{\max} =$ \unit[12]{\r{A}}) at least once over the simulation time (normalized to mean 0 and standard deviation 1) serve as input features of the corresponding neural network
of the ISOKANN algorithm \cite{ISOKANNjl}. 
The action of the Koopman operator in (\ref{equ:power}) is estimated with one sample each forward and backward along the time axis, justified by the reversibility of the system. This is used to train the $\chi$-function. With taking forward and backward steps, it is avoided that $\chi$ has all its mass in the terminal point. An uni-directed trajectory, \corrected{i.e. if only forward or only backward steps were taken}, would ``transport'' the $\chi$ values along the trajectory to the final point in the estimation of the Koopman operator: \corrected{The Koopman estimation of each point in \eqref{equ:power} would simply be the $\chi$ value at the trajectories predecessors point. Over time all points would attain the value of the first point, whereas the shift-scale would enforce the last point to remain distinct.} 
%\corrected{The reason is that the $\chi$ values would only be passed in one direction with each iteration step at each sample. They would be passed on to their neighbours step by step until they reached the end.}

%Using just a single forward-trajectory inhibits convergence, but may nevertheless "get stuck" at the slower parts of the process, such that early stopping may identify the slowest subproces. 
%When using the time-reversed data as well we expect convergence to a unique solution, which however is then dominated by the start and end point of the trajectory.

The neural network is a multilayer perceptron with 3 fully connected hidden layers of size $(6161,336,18)$ with sigmoid activation functions and a single linear output neuron.
For the optimisation we use ADAM with a learning-rate of $\eta=1\mathrm{e}{-4}$ with a $L2$ regularization of $\lambda = 1\mathrm{e}{-2}$ and a minibatch-size of 128. After training $\chi$ for 30,000 iterations \corrected{we concluded convergence based on the plateauing of the mean squared error of the ISOKANN residuals at around $5e-4$}. The described shortest path is extracted with $\sigma = 0.7$ resulting in 1,231 selected frames.

Using a large regularization for the neural network enforces a smooth structure to the $\chi$ function which may also be understood as a smoothing of the data and thereby artificially connecting spatially adjacent samples. With regularization ISOKANN, thus, isolates the spatially large transitions which amount to macroscopic changes.

\subsection{Application to MOR}
\begin{figure}[t]
    \centering
    \includegraphics[width=0.8\linewidth]{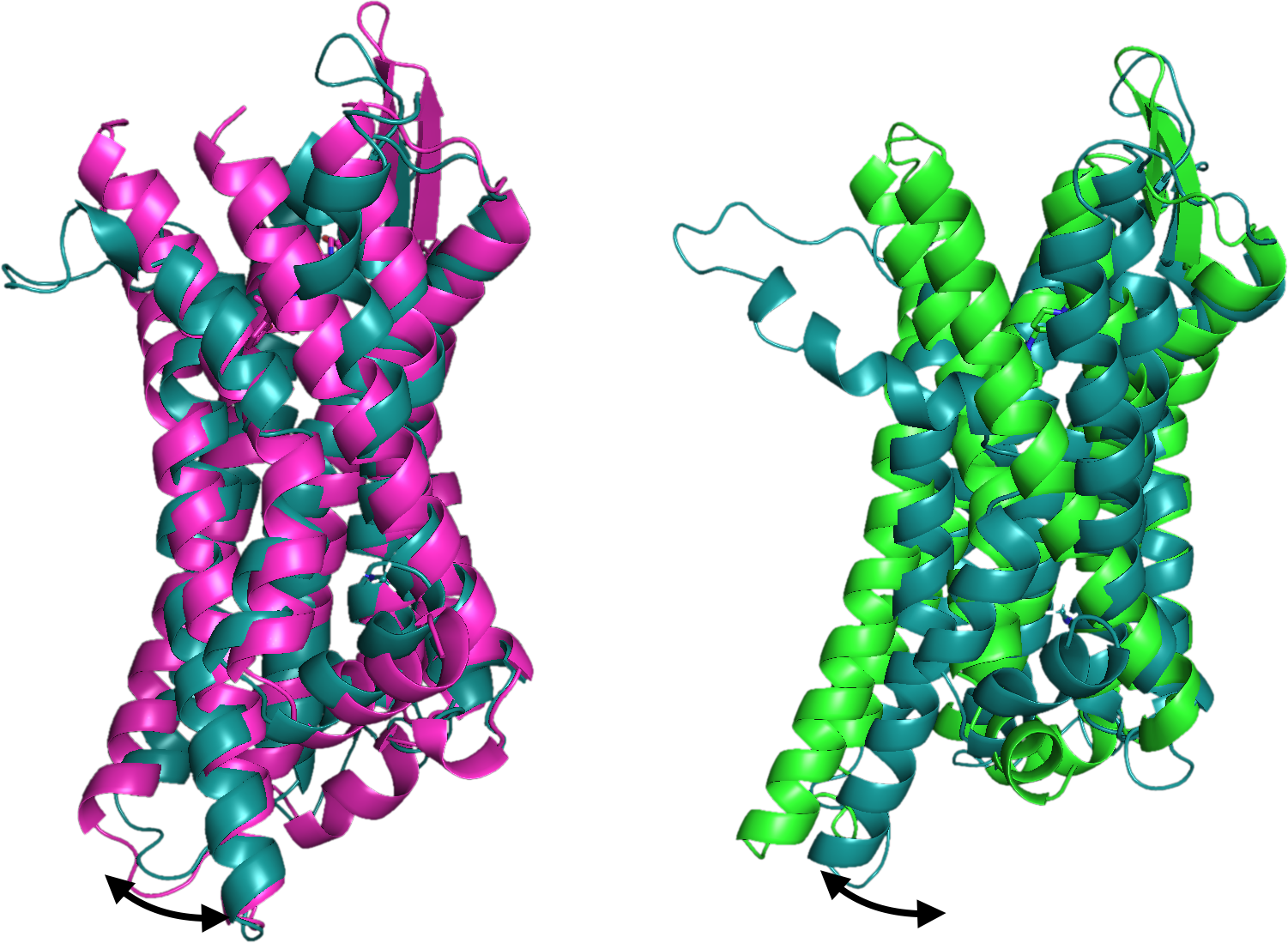}
    \caption{
    %Relevant results are obtained from ISOKANN: The characterized ``orange'' path in Fig.~\ref{fig:chi_along_time} shows that the simulated MOR moves from an inactive state of MOR (purple) to an active state of MOR (green) as a result of disulfide bond between amino acid Cysteine 159 and 251 due to inflammatory environment. The white arrow indicates the movement of transmembrane helix 6 taking place along the path. We show a superposition of the found structures with known structures from literature (gray, PDB id: 7UL4 left and 8EF5 right).
    Visualization of the MOR states (teal) at the beginning (left) and end (right) of the transition path from figure Fig.~\ref{fig:chi_along_time} (orange). The superimposed purple and green structures are known to be representative of the inactive (PDB 7UL4) resp. active (PDB 8EF5) states\corrected{}[(PDB 7UL4 and 8EF5)]. We can see that ISOKANN isolates close similar to these known metastabilities from the simulation trajectories without their a priori knowledge. \corrected{The reactive path clearly displays the tilting motion (black arrow) undergone by the transmembrane helix 6.}}
    \label{fig:active_inactive}
\end{figure}

ISOKANN serves as an efficient tool to analyse rare events in the simulation of the $\mu$-opioid receptor. As mentioned earlier, finding a solution function $\chi$ of (\ref{eq:isosolution}) with a low exit rate $c_1$ corresponds to the identification of a ``reaction coordinate''. 
In Fig.~\ref{fig:chi_along_time} the resulting $\chi$-values along 10 independent molecular simulations are shown (blue dots). One can see that the lowest and highest values of $\chi$ are not to be found within one trajectory. Using ISOKANN, it is possible to extract the ``\corrected{dynamically}[temporal] most distant'' frames from the simulation, see Fig.~\ref{fig:active_inactive}. Indeed the $\chi$-values correspond to a reaction coordinate for the transition from an inactive to an active macro state of MOR. Although none of the 10 trajectories simulates this process completely, we can extract the time-determining steps along the reaction path by using the shortest path routine described in section \ref{chap:methodik}. The picked path is rather small in length (1,231 frames, orange in Fig.~\ref{fig:chi_along_time}) as compared to entire trajectory of 10,000 stored frames. \corrected{It displays a physically plausible and compellingly smooth transition between the extremal states}.

\corrected{Whereas it was not at all clear whether the supplied simulation data was sufficient a priori, the resulting path identifies the known crystal structures and a path between them.} Fig. \ref{fig:active_inactive} displays the modified MOR (teal) with a disulfide bond at the starting point of the reaction path aligning with the inactive structure of the unmodified, natural state of the MOR (purple). \corrected{Laboratory} experiments suggest partial activation of the MOR which is confirmed by the shortest path analysis in the tilting movement of the transmembrane helix 6 (bottom left) outwards like it is seen in the fully activated, natural MOR crystal structure (green).
It's noteworthy that the start and end points of the reactive path were determined without prior knowledge of the crystal structures. 
%Note that both start and end points of the reactive path where determined without any prior knowledge about the crystal structures.
\corrected{Extracting these extremal states and the transition path from the raw trajectory data frame-by-frame would pose a considerable challenge.}

Using a linear regression to estimate $\gamma_1, \gamma_2$ in (\ref{eq:isosolution}) we can compute the exit rate for a given lag time $\tau=$ \unit[0.1]{ns} by \eqref{equ:16} resulting in $c_1 \approx$ \unit[0.06]{ns$^{-1}$}.

\section{Conclusion}
%In order to transfer the concept of mean holding probabilities from a micro to a macro point of view, the equation $p_\chi(\statex, t)=\chi(\statex) \mathrm{e}^{-c_1 t}$ has been introduced.
In bridging the conceptual gap between microstate and macrostate analyses, we introduced the notion of mean holding probabilities $p_\chi(\boldsymbol{x}, t)=\chi(\boldsymbol{x}) \mathrm{e}^{-c_1 t}$ represented in terms of membership functions $\chi$ which we interpret as the macrostate itself.
Taking this as a theoretical starting point, the computation of mean holding times $t_{mh}\propto\chi$, which generalize their classical set-based definitions, and of reaction paths along $\boldsymbol{r}\propto -\nabla\chi$ has been shown. Being proportional to a mean holding time, $\chi(\statex)$ represents a kind of temporal order of micro states $\statex \in \Omega$ in a high-dimensional space $\Omega$ thereby serving as a reaction coordinate.

We briefly described ISOKANN, a machine-learning-based algorithm, which we use to approximately solve the high-dimensional partial differential equation (\ref{eq:Lchi}) defining $\chi$.
In a further step we interpret the values of the obtained solution $\chi$ to  define the weights of edges of a graph, in which the vertices represent biomolecular micro states of an opioid-receptor simulation. Solving a shortest path problem for this graph allows us to obtain a subsample of the simulation data which captures the time-determining steps of macro molecular transitions.  

Our method is able to pick micro states from different independent MD trajectories generated for the same biomolecular system in order to combine them into one ``temporally and spatially'' ordered path between macro states, see Fig.~\ref{fig:chi_along_time}. This path shows the time-determining steps of a rare transition event between the macro states.

In our example, this approach effectively condenses 10,000 frames into a concise set of 1,231 frames. Using ISOKANN together with shortest path computation extracts what ``really is to be seen'' in the trajectories. In this case, the extracted path accurately depicts the transition of the modified MOR from an inactive to an active state as also indicated in experimental conditions. It further enhances our understanding of the role of a disulfide bond resulting from oxidative stress. Supported by this findings detailed experiments highlighting the role of implicated cysteins pair (159 and 251) out of other 2 pairs are planned. This example highlights ISOKANN's potential to significantly streamline the analysis of long-term MD simulations and extract meaningful reaction paths.

\acknowledgement{
 The research of A.~Sikorski was funded by the DFG through the CRC 1114 ``Scaling Cascades in Complex Systems'' (project B03). The research of R.~J.~Rabben was funded by the NHR Graduate School of the NHR Alliance. The research of S.~Chewle was funded by the BMBF through the project CCMAI (funding code 01GQ2109A). 
}

%\backmatter
%\bibliographystyle{plain}
%\printbibliography[env=bibnumeric]

%%%%%%%%%%%%%%%%%%%%%%%%%%%%%%%%%%%%%

\bibliographystyle{abbrv}
\bibliography{literature}

\begin{thebibliography}{10}

\bibitem{darcq2018opioid}
E.~Darcq and B.~L. Kieffer.
\newblock Opioid receptors: drivers to addiction?
\newblock {\em Nature Reviews Neuroscience}, 19(8):499--514, 2018.

\bibitem{pcca-plus_grundpaper}
P.~Deuflhard and M.~Weber.
\newblock {Robust Perron Cluster Analysis in Conformation Dynamics}.
\newblock {\em Linear Algebra and its Applications}, 398c:161--184, 2005.

\bibitem{feynman-kac_paper_weber}
N.~Ernst, K.~Fackeldey, A.~Volkamer, O.~Opatz, and M.~Weber.
\newblock Computation of temperature-dependent dissociation rates of metastable protein--ligand complexes.
\newblock {\em Molecular Simulation}, 45(11):904--911, 2019.

\bibitem{bellmanford}
L.~R. Ford.
\newblock Network flow theory.
\newblock {\em Rand Corporation Santa Monica, CA}, 1956.

\bibitem{feynman-kac_paper_original}
H.~Gzyl.
\newblock {The Feynman-Kac formula and the Hamilton-Jacobi equation}.
\newblock {\em Journal of Mathematical Analysis and Applications}, 142(1):74--82, 1989.

\bibitem{Leimkuhler.2015}
B.~Leimkuhler and C.~Matthews.
\newblock {\em Molecular Dynamics: With Deterministic and Stochastic Numerical Methods}, volume~39 of {\em Interdisciplinary Applied Mathematics}.
\newblock {Springer International Publishing} and {Imprint: Springer}, Cham, 1st ed. 2015 edition, 2015.

\bibitem{Mura_2014}
C.~Mura and C.~E. McAnany.
\newblock An introduction to biomolecular simulations and docking.
\newblock {\em Molecular Simulation}, 40(10–11):732–764, Aug. 2014.

\bibitem{Norris.1998}
J.~R. Norris.
\newblock {\em Markov chains}, volume~2 of {\em Cambridge series on statistical and probabilistic mathematics}.
\newblock {Cambridge University Press}, Cambridge, 1998.

\bibitem{pavliotis}
G.~A. Pavliotis.
\newblock {\em Stochastic Processes and Applications -- Diffusion Processes, the Fokker-Planck and Langevin Equations}.
\newblock Springer, 2014.

\bibitem{pert1973opiate}
C.~B. Pert and S.~H. Snyder.
\newblock Opiate receptor: demonstration in nervous tissue.
\newblock {\em Science}, 179(4077):1011--1014, 1973.

\bibitem{Prinz}
J.~Prinz, H.~Wu, M.~Sarich, B.~Keller, M.~Senne, M.~Held, J.~Chodera, C.~Sch\"utte, and F.~No\'e.
\newblock {Markov models of molecular kinetics: Generation and validation}.
\newblock {\em J Chem Phys.}, 134(17):174105, 2011.

\bibitem{isokann_paper}
R.~J. Rabben, S.~Ray, and M.~Weber.
\newblock {ISOKANN: Invariant subspaces of Koopman operators learned by a neural network}.
\newblock {\em The Journal of Chemical Physics}, 153(11):114109, 2020.

\bibitem{reeh1996tissue}
P.~W. Reeh and K.~H. Steen.
\newblock Tissue acidosis in nociception and pain.
\newblock {\em Progress in brain research}, 113:143--151, 1996.

\bibitem{robertson2022structure}
M.~J. Robertson, M.~M. Papasergi-Scott, F.~He, A.~B. Seven, J.~G. Meyerowitz, O.~Panova, M.~C. Peroto, T.~Che, and G.~Skiniotis.
\newblock Structure determination of inactive-state gpcrs with a universal nanobody.
\newblock {\em Nature Structural \& Molecular Biology}, 29(12):1188--1195, 2022.

\bibitem{schuette}
C.~Schütte, S.~Klus, and C.~Hartmann.
\newblock Overcoming the timescale barrier in molecular dynamics: Transfer operators, variational principles and machine learning.
\newblock {\em Acta Numerica}, 32:517–673, 2023.

\bibitem{ISOKANNjl}
A.~Sikorski.
\newblock {Julia Package: ISOKANN.jl}.
\newblock \url{https://github.com/axsk/ISOKANN.jl}, 2023.

\bibitem{sikribweb}
A.~Sikorski, E.~Ribera~Borrell, and M.~Weber.
\newblock {Learning Koopman eigenfunctions of stochastic diffusions with optimal importance sampling and ISOKANN}.
\newblock {\em Journal of Mathematical Physics}, 65(1):013502, 01 2024.

\bibitem{nontoxic_pain_killer}
V.~Spahn, G.~{Del Vecchio}, D.~Labuz, A.~Rodriguez-Gaztelumendi, N.~Massaly, J.~Temp, V.~Durmaz, P.~Sabri, M.~Reidelbach, H.~Machelska, M.~Weber, and C.~Stein.
\newblock A nontoxic pain killer designed by modeling of pathological receptor conformations.
\newblock {\em Science (New York, N.Y.)}, 355(6328):966--969, 2017.

\bibitem{habil}
M.~Weber.
\newblock {\em {A Subspace Approach to Molecular Markov State Models via a New Infinitesimal Generator}}.
\newblock Habilitation thesis, FU Berlin, 2011.

\bibitem{ernst}
M.~Weber and N.~Ernst.
\newblock A fuzzy-set theoretical framework for computing exit rates of rare events in potential-driven diffusion processes, 2017.

\bibitem{2008_weber_kube}
M.~{Weber} and S.~{Kube}.
\newblock {Preserving the Markov Property of Reduced Reversible Markov Chains}.
\newblock In T.~E. {Simos} and C.~{Tsitouras}, editors, {\em Numerical Analysis and Applied Mathematics: International Conference on Numerical Analysis and Applied Mathematics 2008}, volume 1048 of {\em American Institute of Physics Conference Series}, pages 593--596, Sept. 2008.

\bibitem{zhuang2022molecular}
Y.~Zhuang, Y.~Wang, B.~He, X.~He, X.~E. Zhou, S.~Guo, Q.~Rao, J.~Yang, J.~Liu, Q.~Zhou, et~al.
\newblock Molecular recognition of morphine and fentanyl by the human $\mu$-opioid receptor.
\newblock {\em Cell}, 185(23):4361--4375, 2022.

\end{thebibliography}

%%%%%%%%%%%%%%%%%%%%%%%%%%%%%%%%%%%%%
\end{document}